\documentclass[nofootinbib]{revtex4}
\usepackage{graphicx}
\usepackage{fancyhdr}
\usepackage{amsmath}
\usepackage{color}

\begin{document}

\title{Bare Higgs mass and potential at ultraviolet cutoff}

%

\author{Yuta Hamada}
\author{Hikaru Kawai}
\affiliation{Department of Physics, Kyoto University, Kyoto 606-8502, Japan}
\author{Kin-ya Oda}
\affiliation{Department of Physics, Osaka University, Osaka 560-0043, Japan}

\begin{abstract}
We first review the current status of the top mass determination paying attention to the difference between the $\overline{\text{MS}}$ and pole masses.
Then we present our recent result on the bare Higgs mass at a very high ultraviolet cutoff scale.
\end{abstract}

\maketitle

\thispagestyle{fancy}


\section{Introduction}
It is more and more likely that the 125\,GeV particle discovered at the Large Hadron Collider (LHC)~\cite{Aad:2012tfa,Chatrchyan:2012ufa} is the Standard Model (SM) Higgs. Its couplings to the $W$ and $Z$ gauge bosons, to the top and bottom quarks, and to the tau lepton are all consistent to those in the SM within one standard deviation even though their values vary two orders of magnitude, see e.g.\ Ref.~\cite{Giardino:2013bma}.
No hint of new physics beyond the SM has been found so far at the LHC up to 1\,TeV.
It is important to examine to what scale the SM can be a valid effective description of nature.

In Ref.~\cite{Froggatt:1995rt}, Froggatt and Nielsen have predicted the top and Higgs masses to be $173\pm5\,\text{GeV}$ and $135\pm9\,\text{GeV}$, respectively, based on the assumption that the SM Higgs potential must have another minimum at the Planck scale and that its height is order-of-magnitude-wise degenerate to the SM one. (This assumption is equivalent to the vanishing Higgs quartic coupling and its beta function at the Planck scale.) The success of this prediction indicates that at least the top-Higgs sector of the SM is not much altered up to a very high ultraviolet (UV) cutoff scale.

As all the parameters in the SM are fixed by the Higgs mass determination, we can now obtain the \emph{bare} parameters at the UV cutoff scale, which then become important inputs for a given UV completion of the SM. If the UV theory fails to fit them, it is killed. 

The parameters in the SM are dimensionless except for the Higgs mass (or equivalently its vacuum expectation value (VEV)). The dimensionless bare coupling constants can be approximated by the running ones at the UV cutoff scale, see e.g.\ Appendix of Ref.~\cite{Hamada:2012bp}. The latter can be evaluated through the standard renormalization group equations (RGEs) once the low energy inputs are given. After fixing all the dimensionless bare couplings, the last remaining one is the bare Higgs mass which is the main subject of this work.

\section{Top quark Yukawa coupling at top mass scale}
The largest ambiguity for the Higgs mass parameter is coming from the low energy input of the top Yukawa coupling at the top mass scale.
Let us review the present status of its determination.
Important point is the distinction between the modified minimal subtraction ($\overline{\text{MS}}$) mass and the pole one, which are utilized, respectively, in the $\overline{\text{MS}}$ and on-shell schemes.

Currently the most accurate value of the top quark mass is obtained from combination of the Tevatron data, basically reconstructed as an invariant mass of its decay products~\cite{CDF:2013jga}:
\begin{align}
M_t^\text{Tevatron}	&=	173.20\pm0.87\,\text{GeV},
	\label{Tevatron mass}
\end{align}
whereas the similar analysis of the LHC data gives~\cite{LHC top mass,Deliot:2013hc}: $M_t^\text{LHC}=173.3\pm1.4\,\text{GeV}$.
(If we naively combine these two results, we get
$M_t^\text{inv}=173.2\pm0.7\,\text{GeV}$,
which is of 0.4\% accuracy.)
However, the authors of Ref.~\cite{Alekhin:2012py} criticize that the top quark mass, measured at the Tevatron and LHC via kinematical reconstruction from the top quark decay products and comparison to Monte Carlo simulations, is not necessarily the pole mass $M_t$ but is merely the mass parameter in the Monte Carlo program which does not resort to any given renormalization scheme. The point is that the mass of the colored top quark is reconstructed from the color singlet final states.\footnote{
Note however that the pole mass of the colored quark is well defined to all orders in perturbation theory and that its infrared renormalon ambiguity appears only at the non-perturbative level of order $\Lambda_\text{QCD}$.
}
To circumvent this problem, they propose to determine the $\overline{\text{MS}}$ top quark mass directly from the dependence of the inclusive $t\bar t$ cross section on it. In Ref.~\cite{Alekhin:2012py}, the observed values at Tevatron:
\begin{align}
\sigma(p\bar p\to t\bar t+X)
	&=	7.56^{+0.63}_{-0.56}\,\text{pb}\quad\text{(D0)}
		\qquad\text{and}\qquad
		7.50^{+0.48}_{-0.48}\,\text{pb}\quad\text{(CDF)}
\end{align}
are combined and fit by the theoretical prediction, which is obtained by using four different parton distribution functions (PDFs) at the NNLO and by including the NNLO QCD contributions to $\sigma(p\bar p\to t\bar t+X)$. The resultant value of the $\overline{\text{MS}}$ running top mass at the top mass scale becomes~\cite{Alekhin:2012py}:
\begin{align}
m_t^\text{QCD}(M_t)
	&=	163.3\pm2.7\,\text{GeV}.
		\label{MSbar top mass}
\end{align}
In the above computation, the NLO electroweak (EW) radiative corrections ($\propto\alpha\alpha_s$) to $\sigma(p\bar p\to t\bar t+X)$ are neglected. The ratio of the sum of such EW corrections to the $t\bar t$ total cross section at the Tevatron in the on-shell scheme is shown to be less than 0.2\% for the Higgs mass 120--200\,GeV and the top pole mass 165--180\,GeV~\cite{Kuhn:2006vh}. 

Theoretically the $\overline{\text{MS}}$ mass in Eq.~\eqref{MSbar top mass} is related to the pole mass $M_t$ by
\begin{align}
m_t^\text{QCD}(M_t)
	&=	M_t\left(
			1+\delta_{t}^\text{QCD}(M_t)\right),
			\label{Djouadi's MS-bar mass}
\end{align}
where up to the NNLO QCD corrections of $O(\alpha_s^3)$, see e.g.\ Ref.~\cite{Jegerlehner:2012kn},
\begin{align}
\delta_{t}^\text{QCD}(M_t)
	&=		-{4\over3}{\alpha_s(M_t)\over\pi}
			-9.125\left(\alpha_s(M_t)\over\pi\right)^2
			-80.405\left(\alpha_s(M_t)\over\pi\right)^3,
\end{align}
with $\alpha_s(\mu)=g_s^2(\mu)/4\pi$ being the strong coupling in the six flavor $\overline{\text{MS}}$ scheme.
This relation result in the pole mass~\cite{Alekhin:2012py}
\begin{align}
M_t	&=	173.3\pm2.8\,\text{GeV}.
	\label{Djouadi pole mass}
\end{align}
In Ref.~\cite{Alekhin:2012py} the pole mass of the top quark $M_t$ is also directly extracted from the NNLO theory prediction using the on-shell scheme. The resultant central value varies $169.9$--$172.7$\,GeV, depending on the PDF, with the error less than 2.4\,GeV for each. These values are consistent to Eq.~\eqref{Djouadi pole mass}. To summarize, the derived pole mass is close to the experimentally obtained invariant mass~\eqref{Tevatron mass}.

It is customary to define the $\overline{\text{MS}}$ running VEV $v(\mu)$ in such a way that
\begin{align}
-m^2(\mu)=\lambda(\mu)\,v^2(\mu)
	\label{running VEV definition}
\end{align}
holds for the potential
\begin{align}
\mathcal V=m^2\phi^\dagger\phi+\lambda\left(\phi^\dagger\phi\right)^2,
\end{align}
with $\left\langle\phi\right\rangle=v/\sqrt{2}$.
Then the $\overline{\text{MS}}$ top mass is commonly defined as, see e.g.\ Ref.~\cite{Jegerlehner:2012kn},
\begin{align}
m_t(\mu)
	&=	{y_t(\mu)\,v(\mu)\over\sqrt{2}}.
		\label{MSbar relation}
\end{align}
This definition of the $\overline{\text{MS}}$ mass leads to~\cite{Jegerlehner:2012kn}
\begin{align}
m_t(M_t)
	&=	m_t^\text{QCD}(M_t)
		+M_t\,\Delta_t^\text{EW}(M_t), \label{MSbar and pole masses}
\end{align}
where, taking into account up to NLO EW contributions of $O(\alpha\alpha_s)$,
\begin{align}
\Delta_t^\text{EW}(M_t)
	&=		0.0664-0.00115\left({M_H\over\text{GeV}}-125\right).
			\label{EW NLO corrections}
\end{align}
This discrepancy~\eqref{EW NLO corrections} is due to the definition of the $\overline{\text{MS}}$ top mass via Eq.~\eqref{MSbar relation}, and is dominantly coming from the tadpole contribution to the shift of $v(\mu)$. If we instead use the definition of the $\overline{\text{MS}}$ mass $m_t^\text{QCD}(\mu)=y_t(\mu)V/\sqrt{2}$ with $V=\left(\sqrt{2}G_\mu\right)^{-1/2}=246.22$\,GeV, where $G_\mu=1.1663787(6)\times10^{-5}\,\text{GeV}^{-2}$ is the Fermi constant determined from the muon life time, then we would get the one given in Eq.~\eqref{Djouadi's MS-bar mass}. 

Plugging~\cite{Degrassi:2012ry}
\begin{align}
g_s(M_t)
	&=	1.1645
		+0.0031\left({\alpha_s(M_Z)-0.1184\over0.0007}\right)
		-0.00046\left({M_t\over\text{GeV}}-173.15\right)
\end{align}
into Eq.~\eqref{MSbar and pole masses}, we get
\begin{align}
m_t(M_t)
	&=	M_t\left[
			1.00658
			-0.00041\left({\alpha_s(M_Z)-0.1184\over0.0007}\right)
			+0.00006\left({M_t\over\text{GeV}}-173.15\right)
			-0.00115\left({M_H\over\text{GeV}}-125\right)
			\right]. 
			\label{mtMt}
\end{align}
The $\overline{\text{MS}}$ mass~\eqref{MSbar and pole masses} becomes \emph{larger} than the top quark pole mass $M_t$~\cite{Jegerlehner:2012kn}. 

Let us review the derivation of $\Delta_t^\text{EW}(M_t)$ more in detail. First the $\overline{\text{MS}}$ running VEV $v(\mu)$ is obtained and then it is multiplied to the $\overline{\text{MS}}$ running Yukawa $y_t(\mu)$ in order to obtain the running mass $m_t(\mu)$ in Eq.~\eqref{MSbar and pole masses}.
$v(\mu)$ can be read from the Fermi constant $G_F(\mu)=1/\sqrt{2}v^2(\mu)$:
\begin{align}
G_\mu
	&=	G_F(\mu)\left(1+\Delta_{G_F,\alpha}+\Delta_{G_F,\alpha\alpha_s}+\cdots\right)
	=	G_F(\mu)\left[1+{\alpha_2(\mu)\over4\pi}{m_t^4(\mu)\over m_W^2(\mu)\,m_H^2(\mu)}\left(6-12\ln{m_t(\mu)\over\mu}\right)+\cdots\right].
		\label{running Fermi constant}
\end{align}
The $O(\alpha)$ and $O(\alpha\alpha_s)$ contributions $\Delta_{G_F,\alpha}$ and $\Delta_{G_F,\alpha\alpha_s}$ are given in Eqs.~(A.3) and (A.6) in Ref.~\cite{Bezrukov:2012sa}. The dominant tadpole contribution is picked up in the last step in Eq.~\eqref{running Fermi constant} for explicitness. The resultant $\overline{\text{MS}}$ VEV is $v(M_t)\sim 260$\,GeV at the top mass scale. On the other hand, the $\overline{\text{MS}}$ Yukawa coupling is given by~\cite{Hempfling:1994ar}
\begin{align}
y_t(\mu)
	&=	{M_t\over\sqrt{2}V}\left(
			1
			+\delta_t^\text{QCD}(\mu)
			+\delta_{t,\alpha}^\text{QED}(\mu)
			+\delta_{t,\alpha}^W(\mu)
			+\delta_{t,\alpha\alpha_s}^\text{EW}(\mu)
			+\cdots\right),
			\label{MS-bar Yukawa}
\end{align}
where the $O(\alpha)$ corrections are
\begin{align}
\delta_{t,\alpha}^\text{QED}(\mu)
	&=	{Q_t^2\alpha(\mu)\over4\pi}\left(6\log{M_t\over\mu}-4\right),\\
\delta_{t,\alpha}^W(\mu)
	&=	{G_\mu\,m_t^2(\mu)\over8\pi^2\sqrt{2}}\left[
			-\left(2N_c+3\right)\ln{M_t\over\mu}+{N_c\over2}+4-r+2r\left(2r-3\right)\ln(4r)
			-8r^2\left({1\over r}-1\right)^{3/2}\arccos\sqrt{r}
			\right],
\end{align}
with $Q_t=2/3$, $N_c=3$, and $r=M_H^2/4M_t^2$. 
The resultant explicit analytic formula of $\Delta_t^\text{EW}(M_t)$ is given in Ref.~\cite{Jegerlehner:2003py} that takes into account up to the NLO EW corrections of $O(\alpha\alpha_s)$.
The tiny $O(\alpha\alpha_s)$ correction to the Yukawa coupling, $\delta_{t,\alpha\alpha_s}^\text{EW}(\mu)$ in Eq.~\eqref{MS-bar Yukawa}, can be read off from $\Delta_{t,\alpha\alpha_s}^\text{EW}$ by subtracting the tadpole contribution to $v(\mu)$.
In Ref.~\cite{Degrassi:2012ry}, numerical value of Eq.~\eqref{MS-bar Yukawa} is evaluated as
\begin{align}
y_t(M_t)
	&=	0.93587
		+0.00557\left({M_t\over\text{GeV}}-173.15\right)
		-0.00003\left({M_H\over\text{GeV}}-125\right)
		-0.00041\left(\alpha_s(M_Z)-0.1184\over0.0007\right)
		\pm0.00200_\text{th}.
		\label{Strumia Yukawa}
\end{align}
Multiplying Eq.~\eqref{Strumia Yukawa} by the running VEV $v(\mu)$ that is read from Eq.~\eqref{running Fermi constant}, we obtain $\Delta_t^\text{EW}$, and hence the running mass~\eqref{mtMt}.

\section{Bare parameters at high scale}
\begin{figure}[tn]
\begin{center}
\includegraphics[width=.5\textwidth]{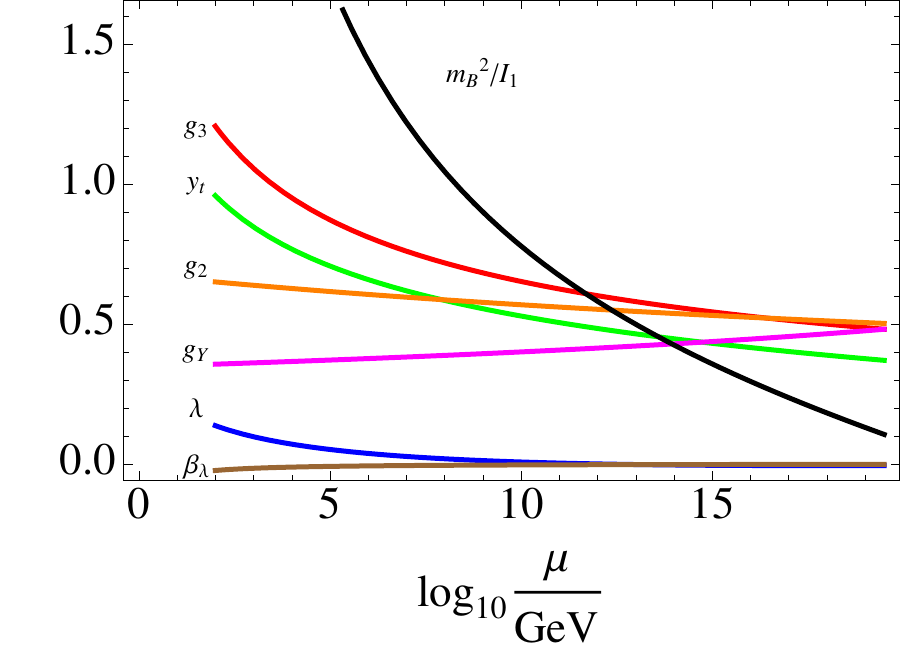}
\caption{RGE running of the SM couplings and of the beta function for the quartic coupling $\beta_\lambda$, except for $m_B^2$ which is \emph{not} a running mass but is the (quadratically divergent part of) bare Higgs mass parameter, obtained by taking each scale on the horizontal axis to be the cutoff $\Lambda$. $I_1=\Lambda^2/16\pi^2$ is the one loop integral.}\label{SMcouplings}
\end{center}
\end{figure}

In this proceedings we show our result~\cite{Hamada:2012bp} based on Eq.~\eqref{Strumia Yukawa} with the pole mass~\eqref{Djouadi pole mass}. 
In Figure~\ref{SMcouplings}, we show a plot with the two loop RGEs, summarized in Ref.~\cite{Hamada:2012bp}, for the dimensionless SM couplings, with the low energy boundary condition for the top Yukawa as explained above, namely with $y_t(M_t)=0.93587$. $\beta_\lambda$ is the beta function for the Higgs quartic coupling: $\beta_\lambda=d\lambda/d\ln\mu$. $m_B^2/I_1$ is explained in the following.

In the bare perturbation theory, the renormalized Higgs mass squared parameter is given at the one loop level by
\begin{align}
m_R^2
	&=	m_{B,\text{1-loop}}^2+\left(6\lambda_B+\frac{3}{4} g_{YB}^2+\frac{9}{4} g_{2B}^2-6y_{tB}^2\right)I_1+\delta m^2,
		\label{one-loop renormalized mass}
\end{align}
where $I_1=\int^\Lambda d^4p/(2\pi)^4p^2=\Lambda^2/16\pi^2$ is the quadratically divergent one loop integral over the Euclidean momentum. In obtaining the expression~\eqref{one-loop renormalized mass}, we assume existence of an underlying gauge invariant regularization, such as string theory, and freely shift the integrated momenta. The requirement that the sum in the parentheses in Eq.~\eqref{one-loop renormalized mass} to be zero is the celebrated Veltman condition. In a mass independent renormalization scheme, including the dimensional regularization, the bare mass squared $m_B^2$ is chosen in such a way that the renormalized mass parameter becomes zero, $m_R^2=0$, when $\delta m^2=0$; and then non-zero $\delta m^2$ is introduced as a perturbation. (This choice of the bare mass $m_B^2$ to cancel the quadratic divergence is automatic in the dimensional regularization scheme.) Consequently the bare mass $m_B^2$ contains a quadratic divergence: $\Lambda^2$, whereas the running mass $\delta m^2$ only logarithmic one: $\log\Lambda$.
Note that this cancellation of $\Lambda^2$ by $m_B^2$ is done once and for all, and then we never see $\Lambda^2$. Bardeen has argued that therefore the quadratic divergence is not a real problem, see e.g.\ Ref.~\cite{Iso:2013aqa} for a recent review. However, e.g.\ in obtaining a low energy effective field theory from string theory, this procedure of matching $\Lambda^2$ by $m_B^2$ is not a fake, and we focus on this largest part $m_B^2$ in the bare Lagrangian, neglecting the subleasing $\delta m^2\propto v^2\log\Lambda$.

In Ref.~\cite{Hamada:2012bp} we have obtained the bare Higgs mass at two loop orders in the bare perturbation theory:
\begin{align}
m_{B,\,\text{2-loop}}^2
	&=	-\bigg\{
			9y_{tB}^4
			+y_{tB}^2\left(-\frac{7}{12}g_{YB}^2+\frac{9}{4}g_{2B}^2-16g_{3B}^2\right)
			+\frac{77}{16} g_{YB}^4+\frac{243}{16} g_{2B}^4
		+\lambda_B\left(-18y_{tB}^2+3 g_{YB}^2+9 g_{2B}^2\right)
		-10\lambda_B^2
		\bigg\}\,I_2,
		\label{bare mass in SM}
\end{align}
which realizes $m_R^2=0$ for $\delta m^2=0$ at this order, where
\begin{align}
I_2	&=	\int{d^4p\over(2\pi)^4}\int{d^4q\over(2\pi)^4}{1\over p^2q^2(p+q)^2}
\end{align}
is the quadratically divergent two loop integral over the Euclidean momenta. As a non-trivial check of the consistency of our treatment, we have confirmed that the coefficients of the infrared divergent integral $\int{d^4p\over(2\pi)^4p^4}\int{d^4q\over(2\pi)^4q^2}$ cancel out in each gauge invariant set of diagrams. We note that our two loop computation is for the quadratically divergent part $m_B^2$ and is irrelevant to the higher loop result $\propto\left(\log\Lambda\right)^n\Lambda^2$, obtained in Ref.~\cite{Einhorn:1992um} and used in Ref.~\cite{Kolda:2000wi,Casas:2004gh}, which would correspond to the effects of the RGE running of the dimensionless couplings in our language.

\begin{figure}[t]
  \begin{center}
  \hfill
  \includegraphics[width=.5\textwidth]{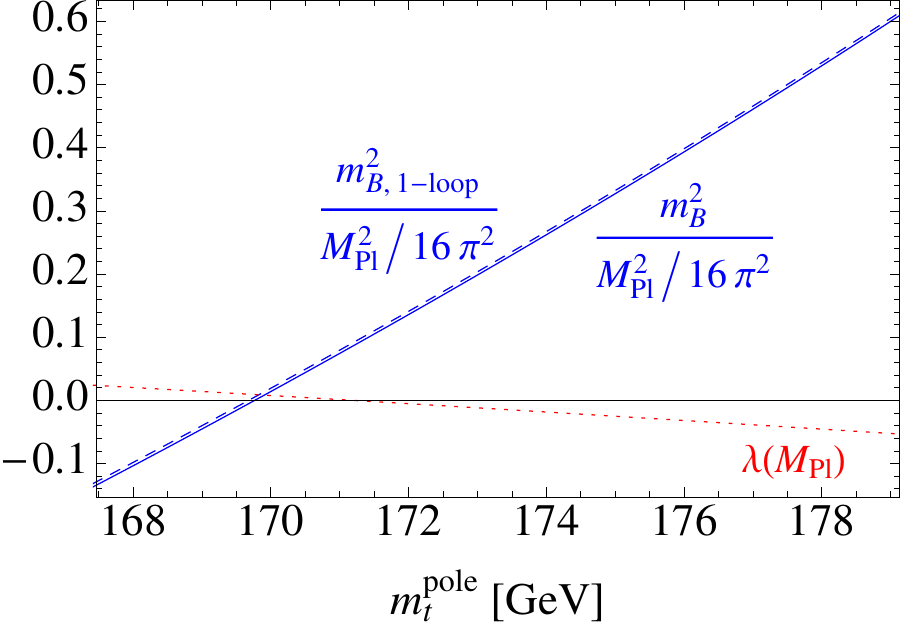}
  \hfill
  \mbox{}
	\end{center}
  \caption{
The blue solid (dashed) line corresponds to the one-plus-two-loop (one-loop) bare mass $m_B^2$ ($m_{B,\,\text{1-loop}}^2$) in units of ${M_\text{Pl}^2/16\pi^2}$ for $\Lambda=M_\text{Pl}$. For comparison, we also plot the quartic coupling $\lambda$ at the Planck scale with the red dotted line. 
}
  \label{fig:msq_vs_mt}
\end{figure}

In obtaining the two-loop corrected bare mass $m_B^2=m_{B,\text{1-loop}}^2+m_{B,\text{2-loop}}^2$, one needs a relation between the one- and two-loop integrals $I_1$ and $I_2$, which is necessarily regularization scheme dependent. When we employ
\begin{align}
\int{d^4p\over p^2}=\int_\epsilon^\infty d\alpha\int d^4p\,e^{-\alpha p^2},
\end{align}
we get
\begin{align}
I_2={I_1\over16\pi^2}\ln{2^6\over3^3}\simeq0.005I_1.
\end{align}
With the blue solid line in Fig.~\eqref{fig:msq_vs_mt}, we plot $m_B^2/I_1$ when the UV cutoff is at the Planck scale, that is, when $I_1=M_P^2/16\pi^2$. The blue dashed line is the same with the 1-loop only bare mass $m_{B,\text{1-loop}}^2/I_1$. We see that the two-loop correction is small, so is the regularization dependence. For comparison, we also plot the quartic coupling $\lambda$ at the Planck scale with the red dotted line. We see that both become small for $M_t\simeq 170$\,GeV.

\section{Summary}
We have computed two loop correction to the quadratically divergent part of the bare SM Higgs mass in the bare perturbation theory. We have found that in generic regularizations the two loop correction to the bare mass is small. Therefore the regularization dependence is not large.
Both the resultant bare Higgs mass and quartic coupling can become small if the top quark pole mass is around 170\,GeV. Possible consequences of this result will be presented elsewhere.

\begin{acknowledgments}
We are grateful to Abdelhak Djouadi, Mikhail Yu.\ Kalmykov, and Bernd A.\ Kniehl for valuable comments.
\end{acknowledgments}

\bigskip 

\end{document}